\documentclass[letterpaper,12pt]{article}
\usepackage{tabularx} 
\usepackage{amsmath}  
\usepackage{graphicx} 
\usepackage[margin=1in,letterpaper]{geometry} 
\usepackage{cite} 
\usepackage[final]{hyperref} 
\hypersetup{
	colorlinks=true,       
	linkcolor=blue,        
	citecolor=blue,        
	filecolor=magenta,     
	urlcolor=blue         
}

\newcommand{\T}{T^{\prime}}
\newcommand{\X}{\xi^\prime}
\begin{document}

\title{Odds for an enlightened rather than barren future}
\author{David Haussler, UC Santa Cruz}
\date{\today}
\maketitle

\begin{abstract}
We are at a stage in our evolution where we do not yet know if we will ever communicate with intelligent beings that have evolved  on other planets, yet we are intelligent and curious enough to wonder about this. We find ourselves wondering about this at the very beginning of a long era in which stellar luminosity warms many planets, and by our best models, continues to provide equally good opportunities for intelligent life to evolve. By simple Bayesian reasoning, if, as we believe, intelligent life forms have the same propensity to evolve later on other planets as we had to evolve on ours, it follows that they will likely not pass through a similar wondering stage in their evolution. This suggests that the future holds some kind of interstellar communication that will serve to inform newly evolved intelligent life forms that they are not alone before they become curious. 
\end{abstract}

\newline

\section{Introduction}

We have learned that we live in a vast universe, perhaps a multiverse, consisting of the relentless march of a wave function ruled by the gentle but firm hand of Schr\"{o}dinger's equation. There are two problems: (1) we can't solve the equation and (2) we don't know which `we' we are, or even if there are other `we's. Yet despite this, our little species is coming to understand a great deal. So where does the mathematics we are gradually mastering suggest that species like us are likely to be heading, toward enlightening communication with other species or solitary self-destruction?  We argue for the former.

If we assume that we as observers are in any special circumstance we like in order to suit our preconceived notions, then all bets are off, and in fact scientific reasoning itself becomes impossible. This is now abundantly evident after centuries of missteps in this regard. Thus, the proper scientific approach is to adopt some kind of Copernican principle to address the second problem, essentially to assume that we, in our moment of observation, are typical among the possible observing `we's in some mathematically well-defined sense. We then perform Bayesian probabilistic reasoning based on that assumption, conditioned on the data we observe. Setting the notion of typicality can be a difficult issue, yet a decision on typicality is essential if we are to properly take into account observation selection effects, such as the ``anthropic bias'' that arises from the fact that already from the get-go, us humans being here to make observations presupposes some conditions.  The most convincing Bayesian arguments made under strong observation selection effects are those whose conclusions are robust to many different versions of the Copernican principle used to define typicality of the observer-moment \cite{bostromBook}. That is, we would like our conclusions to hold under a wide range of reasonable notions of a typical observer making an observation at a typical time. 

Reasoning in a Bayesian fashion from an unassuming notion of a typical observer-moment, what then can science say about the likely future of the human lineage? One extensively discussed argument, known as the Doomsday Argument, concludes that we are likely headed for self-destruction \cite{carter,gott}. This argument invokes one of the simplest kinds of Copernican principles, namely that we, in this case `we' being the individual human pondering the future of our lineage, are {\it a priori} equally likely to be any of the humans past, present and future who might at any moment in their lives do such pondering. Such a set of possible observer-moments is called the {\it reference class}. Under a uniform typicality assumption, all observer-moments in the reference calls are {\it a priori} equally likely. Hence the typical observer-moment is just as likely to be any other in the reference class as it is the one you are experiencing now. Under this uniformity assumption, if you list all the humans in this Doomsday Argument reference class according to their birthdays, then replace each person in this list with their ordered life moments to get a linear  ordering of all observer-moments, our (i.e. your or my) particular observation-moment right now is with 90\% probability not in the first 10\% of that ordering. The inevitable Bayesian consequence of this is that there are unlikely to be more than $10$ times as many human fate-pondering observer-events in the future than there were in the past.  Assuming that only the dead lack curiosity, given that our population was very much smaller and shorter-lived in the past, so that we are now burning through observer-moments at a relatively rapid clip, that does not give us much time until some impending doomsday. 

Many authors have attempted to poke holes in the above Doomsday Argument. What is its greatest weakness? A growing chorus of scientists \cite{bostromBook} would say it is the assumption that ``only the dead lack curiosity.'' Without that assumption, the above Bayesian reasoning merely suggests that there will not be a huge number of humans in the future who will spend time pondering the future fate of humanity the way we do. One can imagine so many scenarios other than self-destruction in which our pondering behavior could be transformed that this is almost tantamount to saying that "probably things will change substantially for humans in the future."  That takes a lot of the ``doom'' out of ``doomsday''. 

We argue that the pondering behavior of intelligent species in the future will change by interstellar communication. In 1950 Enrico Fermi wondered why we haven't yet heard from other intelligent species, or indeed why we haven't been colonized by them \cite{fermiParadox}. It was estimated early on that it would take a sufficiently advanced civilization at most on the order of 100 million years to colonize the galaxy, an estimate that has, if anything, been reinforced by subsequent technological development and analysis. Back then it appeared that, given the approximately 9 billion year age of our galaxy, civilizations should have had time to develop on other planets way ahead of us, and hence communicate with or colonize us. That this has not occurred is the Fermi Paradox. 

The Doomsday Argument has been invoked as a possible solution to the Fermi Paradox. The weakness here is that \textit{all} intelligent species would have to exterminate themselves to resolve the paradox, which seems unlikely \cite{fermiParadox}. Alternate solutions to the Fermi Paradox, such as the suggestion that powerful gamma ray bursts or other large-scale phenomena may periodically sterilize parts of the galaxy of primitive life in a single catastrophe, have the advantage of not requiring many independent extinction events to all occur \cite{annis}. Recent results examining all major sources of radiation and using multiple models suggest that due to the high frequency of such catastrophic evenets, including supernovae, intelligent life was unlikely to have been able to evolve anywhere in the universe much earlier than it did on Earth, providing a cogent resolution to the Fermi Paradox \cite{dayal}.  These catastrophic events are currently subsiding, so we may yet go through Fermi's predicted transition where the galaxy, and perhaps our entire Local Group, is united by interstellar communication and a shared advanced intelligence. Here we provide a Bayesian argument in favor of such a scenario.

There is now abundant evidence for a plethora of other life-habitable planets in our Local Group of galaxies within an, albeit leisurely, 10 million light year communication range of Earth. Further, analysis indicates that during the stelliferous era of our universe in which such planetary systems are formed and die, many stars will shine within our local galaxy cluster for some trillions of years, about 1,000 times longer than the roughly 15 billion years since the big bang \cite{adams}. During this time there will be a shift from fewer larger, short-lived stars, including the highest luminosity stars that don't last long enough to support the evolution of life on their orbiting planets, to many more smaller, longer-lived and life-supportive stars.

Already we see that approximately 75\% of the stars in our Local Group are of the smallest variety, M stars with between 0.5 and 0.08 solar mass \cite{tarter}. This percentage will grow. These so-called "red dwarfs" are more convective, mixing their hydrogen more efficiently for a slower but more durable and complete fusion reaction. While their luminosities are smaller (a star with 0.5 solar mass has luminosity about 6\% that our sun, and a star of 0.1 solar mass has luminosity only 0.5\%), the latest models suggest that their luminosity is nevertheless ``usable energy'' that could power carbon-based life requiring liquid water to evolve on suitable planets orbiting them \cite{tarter}. Despite the fact that overall, star formation is winding down relatively quickly, because the red dwarfs are so numerous and long-lived, and because their luminosities will actually increase over their lives, they will maintain a high level of usable energy production in our Local Group for hundreds of billions of years \cite{laughlin,adams}. 

Moreover, the two largest galaxies within our local group, our Milky Way and Andromeda, will merge early in the stelliferous era, probably within 6 billion years \cite{andromedaMerger}, causing a burst of new star formation and reducing the communication distances between their pre-existing stars by roughly two orders of magnitude, bringing it down to the relatively "short" scale of a few 100,000 light years, i.e. thousands of round trips in each billion year epoch. Taken together, these observational models suggest that life in our Local Group has a relatively long and cozy future ahead. 

Let's consider the reference class of all observer-moments of intelligent observers collected together over all life-cradling planets that have or will exist in our Local Group of galaxies. We will call the living individuals generating this reference class our `local cousins'. Loeb, Batista and Sloan have recently calculated that for a co-moving multi-galactic volume with a constant number of baryons like our Local Group, the likelihood for the evolution of life as a function of time is such that 99.9\% of the life-producing potential lies ahead \cite{loeb}. In other words, we are only 0.1\% into the life-supporting period in our Local Group, when time is appropriately weighted by its ability to support the evolution of life. The high frequency of catastrophic events preceeding the evolution of life on Earth further sharpens this by minimizing the intelligent life-producing potential of the past \cite{dayal}. The Loeb 0.1\% number does depend on the assumption that stars as small as proxima centauri, at 1/8th our sun's mass, can support life, but this has support from recent models \cite{tarter}. Even under the much more conservative assumption that only stars with at least 0.27 the mass of our sun can support life, there is still steadily increasing life-producing potential in our Local Group for about 600 billion years \cite {loeb}. If we make the assumption that catastropic events prevented the emergence of intelligent life earlier than 2 billion years ago, which is conservative in light of \cite{dayal} and our observation of not being contacted to date (Fermi Paradox), then we are at most 0.33\% into the intelligent life-supporting period \cite{loeb}.  

Thus, it appears that we human cousins are asking our cosmic questions at a very early point in the intelligent life-supporting period of our Local Group.
Why so fast out of the gates? Why is such an early species as ourselves the one we find asking about the future of intelligent life? This `earliness mystery', which has bearing on the Doomsday Argument and indirectly on the Fermi Paradox, including discussions about the wisdom of expending effort to establish contact with intelligent life on other planets \cite{yuri}, is what we discuss here. 

We assume that we, humans that query, are typical in the sense of being uniformly chosen from among our usable energy-supplied local cousins. Under these assumptions it would be highly unusual for us to be living during the first 0.1\% or 0.33\% of the total production of energy usable for the evolution of intelligent life in our local group of galaxies.  This situation remains equally unusual regardless of the value of the constants in the Drake equation, i.e. the probability of the evolution of intelligent life on a typical planet. This probability could be arbitrarily small and still, as long as it is non-zero (which it must be because we are here), conditioning on the fact that intelligent life that asks the big questions has evolved at least once, a random instance of such intelligent life is still equally likely to be chosen at any time or place where there is enough usable energy, leaving an earliness mystery for us. 

Further, it is easy to see that a high probability of either self-destruction or periodic sterilization of life from external physical sources such as gamma ray bursts or supernovae does not explain the earliness mystery. In fact, the predicted decrease in frequency of such events deepens the mystery: why are we not one of the later civilizations living comfortably between widely separated catastrophes? 

Regarding doomsday, if all of the intelligent life forms wipe themselves out shortly after beginning to ponder their cosmic fate, it is still equally likely that we would be one of those fools about to do this at any time, so why so damn early? 

We apply Bayesian calculations below in an attempt to address the earliness mystery.   A key issue is the existence of communication between the local cousins. We can't solve the earliness mystery by positing that intelligent life is so rare (e.g. by small constants in the Drake equation) that we are the only intelligent life form that will ever exist in our Local Group. To solve it we must have cousins. If they don't communicate, then why would later-evolved beings not perceive themselves as alone and wonder about the existence of other species just as we do? It is again completely {\it ad hoc} to assume that they are some how `just intrinsically different from us'. Further, as with the Drake equation, the probability of being `just intrinsically different from us' can be arbitrarily close to 1 (without being 1) and it does not take the sting out of the Bayesian argument, given that we are wondering early.  

In short, for any otherwise reasonable theory that features lack of communication between independently evolved life forms,  given our early presence in the stelliferous era, by Bayesian reasoning we find that theory to have a low {\it a posteriori} probability relative to a similar theory that features many communicating life forms. Turning this around, we predict that with high probability there will be communication between the local cousins, and that it will involve a substantial fraction of the intelligent observers, and hence possibly involve descendants of the human lineage. 

So we should build those radio telescopes, at least to listen if not to transmit - the wisdom of transmitting remains unresolved. The Bayesian analysis provides no speculation as to the nature of said communication between species, be it hostile or benign. It predicts merely that it will be `informative', or more generously, `enlightening', such that later in the stelliferous era, independently evolving intelligent life forms will not pass through the curious and apparently alone stage that we find ourselves in now.

\section{Results}

We use the mathematical notation of Shrednicki and Hartle \cite{sh2010}, with the observer-moment terminology of Bostrum \cite{bostromBook} for the Bayesian formulation of the arguments. The players are (1) an objective (i.e. third person, or ``bird's eye view'' \cite{tegmark}) theory $T$ of the universe or multiverse (formally, a range of possible Schr\"{o}dinger equations and initial conditions that describe systems with a given property), (2) the data $D$ that are observed by us, here ``us'' being an intelligent observer living in that system observing data from our first person or ``frog's eye view'' of the world, and (3) an \textit{a priori} “xerographic probability distribution” $\xi$ over a given reference class of possible observer-moments that defines the Copernican-like assumption of typicality of us as observers. The  xerographic probability distribution is implicitly assumed to be zero for any combination entity and moment of observation not in the reference class. 

The central mathematical quantity is \(P(D|T,\xi)\), the probability of the data $D$ we observe, under the assumption that the theory $T$ holds and that the observation is made at a typical observer-moment chosen according to the xerographic probability distribution \(\xi\). This is how the theoretical framework consisting of $T$ and $\xi$ explains data, i.e. how it is a theoretical framework. 

From \(P(D|T,\xi)\) we would like to compute the posterior probability that the theoretical framework is appropriate given the data, i.e. \(P(T,\xi|D)\). Mathematically, the only sensible way to do this is using Bayes rule. Bayes rule is more convenient when deployed as a comparison between two theoretical frameworks, say \((T,\xi)\) and \((\T,\X)\). In this case we compute the posterior odds, \(P(T,\xi|D)/P(\T,\X|D)\), instead of the individual posterior probabilities. The posterior odds directly quantify the relative amount of rational belief we should have in one framework versus another, given the observed data. By Bayes rule the posterior odds between two competing frameworks is
\[\frac{P(T,\xi|D)}{P(\T,\X|D)} = \frac{P(D|T,\xi)}{P(D|\T,\X)} \frac{P(T,\xi)}{P(\T,\X)}\]
where $P(T,\xi)$ is the {\it a priori} probability of the framework $(T,\xi).$

To address the earliness mystery, we will consider two complementary theories. In the theory $T$ the Drake constant is large enough so that relatively quickly, an in particular within the time that the first 0.5\% of the energy usable for the evolution of intelligent life is produced in our local galaxy cluster, many intelligent life forms are expected to evolve independently on different newly forming planets (or other suitable places that we don't yet know about), including enough life forms that are sufficiently long-lived and close to each other to form one or more sustained interstellar communication networks, and further that the resulting communication networks are powerful enough to limit or eliminate the time in which newly evolving species later in the stelliferous era are left alone to wonder about their fate.  We call this ({\it interstellar}) {\it convergence}. In the complementary theory $\T$ convergence either does not ever occur, occurs only briefly and then goes away (perhaps multiple times), or occurs late, say only after 50\% percent of the usable energy is expended. Lack of convergence could be for any of a variety of reasons, e.g. one or more constants in the Drake equation could be too low, or plenty of smart species could evolve but kill themselves off before or shortly after they form interstellar communication networks, or they may persist for a long time but for some reason choose not to communicate, or they might consistently demure from communicating with new independently evolved species during the bulk of the stelliferous era, giving new species plenty of time alone to wonder about their future before being contacted by the interstellar communication network and having their questions answered (kind of a Star Trek-style “prime directive”). 

In both theories $T$ and $\T$, intelligent species evolve independently at a roughly constant rate with respect to available usable energy throughout the stelliferous era. Thus, in both cases the xerographic distribution $\xi$ will be roughly uniform over a reference set consisting of all intelligent observers in our Local Group of galaxies during the stelliferous era at all their fate-questioning moments. 

We now want to evaluate the posterior odds
\[\frac{P(T,\xi|D)}{P(\T,\xi|D)} = \frac{P(D|T,\xi)}{P(D|\T,\xi)} \frac{P(T,\xi)}{P(\T,\xi)}.\]

We may divide the first 50\% of the usable energy production that could have led to the evolution of intelligent life in our Local Group into 100 intervals, each consisting of 0.5\% of the energy production (after a suitable ``warm up period" at the begining to give the first intelligent life time to evolve). The observed data are $D =$ the fact that we are intelligent beings wondering if we are alone and we live in the first of these intervals (a very conservative assertion, given that \cite{loeb} estimates were are actually in the first 0.1\%, and a more conservative estimate gives 0.33\% as discussed above).  Under theory $\T$ our wondering can occur with roughly equal probability in any of the 100 intervals, but under the theory $T$ it can only occur in the first interval because interstellar communication prevents it later. Hence, using the uniform xerographic distribution,

\[\frac{P(D|T,\xi)}{P(D|\T,\xi)} \approx 100.\]
That is, the fact that we observe ourselves asking these cosmic questions during the first interval argues about 100 times stronger for theory like \(T\) in which this must happen, than it does for theory like \(\T\) in which this can happen equally in any interval. 

All that is left is to take into account the prior odds ratio. Under a reasonable conditional independence assumption we can reduce the prior odds ratio \(P(T,\xi)/P(\T,\xi)\) to the simpler ratio \(P(T)/P(\T)\).  Putting these results together, we get the result that the rational odds we should place on theory $T$ versus $\T$, given what we know today, are approximately $100 P(T)/P(\T)$. 

This demonstrates that unless your prior probability favors a physics in which \(\T\) holds more than 100 times over one in which \(T\) holds, you should expect a future in which most species communicate with other species across interstellar distances, are thus enlightened as to their place among other alien civilizations, and carry that enlightenment to other future evolving species. That is, you should except an interstellar convergence of knowledge through communication. 

\section{Discussion}
Can we humans rest easy now concerning a self-made doomsday? Unfortunately we cannot. While arguments given above support early interstellar convergence, among theories of early convergence we may still consider, for example, a theory $T_0$ in which there is rampant self-destruction of intelligent species in the beginning of the stelliferous era and then somehow early convergence brings it to a halt, versus another, $T_1$, in which there is early convergence but self-destruction is rare even early in the stelliferous era. Under the same line of reasoning as above, with the same ``we are here now'' observed data $D$ and uniform xerographic prior $\xi$, we find the posterior odds of $T_0$ vs $T_1$ to be close to 1, that is, our observed data can do little or nothing to distinguish $T_0$ from $T_1$. We are back to our unbiased Doomsday Argument analysis: yes something is likely to change for humans, but we are not sure how, and it behooves us to try to make sure it is not self-destruction. 

It is possible that new science will suggest that our stelliferous era won't actually produce as much usable energy as predicted,  or that for other reasons our prior for theory $\T$, where there is either no or late convergence, will become more than 100 times larger than our prior for a similar theory $T$, where there is early convergence. 
Unless this happens, if denying convergence we must
posit that, unlikely as it may be, the fact that we are here asking these questions so early is merely a fluke.

The `fluke' explanation points out that, like most Bayesian arguments based on random observer moments, the argument given here falls far short of scientific proof. A factor of 100, or even 1,000 in favor of one theory over another is tiny compared what we routinely see for established physical theories versus their rivals. Typically the posterior likelihood ratio of a theory over a weaker rival will start relatively low like this, but as independent evidence piles on in favor of the theory over the rival, the ratio skyrockets, to the point where the weaker theory is thoroughly rejected. That has not happened here for the theory $\T$ of late/no convergence. Much stronger, but perhaps hard-to-obtain evidence would be needed to reject it.    

Theories $T$ and $\T$ do not cover all possibilities. There are a spectrum of theories where convergence happens somewhere between 0.5 percent and 50 percent of the way through the energy production of the  stelliferous era.  These are relatively uninteresting. One might reasonably assume that their prior probabilities are lower than those of $T$ or $\T$, because given the powerful effects of various physical constants etc., convergence is bound to happen fairly fast or not at all. More importantly, the conclusions we arrive at from the Bayesian analysis of such cases are just boring intermediates that yield nothing essentially new anyway.

What about other reference classes or xerographic distributions? We get roughly the same results if in our xerographic distribution we uniformly count species, individuals within species, or moments in which individuals are fate-questioning. So the result is robust in that sense. Also, unless our assumptions for the rate or duration of the evolution of new species on new planetary systems are strongly non-uniform, with the rate decreasing something like exponentially in time, the results will be qualitatively similar. As discussed above, the latter issue may hinge on the suitability of the energy produced by red dwarfs to support the evolution of life. 

We could try to narrow the reference class. However, any reference class that does not include both possible observers from other planets and observers who live in the future would systematically prevent us from reasoning about possible interstellar communication, given that it hasn't happen yet, at least to us, and we would need somebody else to talk to. So since we are looking outside and to the future, there is a good reason not to limit the range of stellar systems we consider to any fewer than the set of all that might possibly communicate with each other over an extended period. 
The importance of including non-human observers cannot be overstated, for without this we essentially get just a very much weakened Doomsday Argument.

What about expanding to much broader references classes? There are many more, possibly infinitely many more, planetary systems in our “bubble universe” that are outside or will move outside of our communication reach during the stelliferous era. Add in the observers of their planetary systems, plus yet more observers from systems in other bubble universes within a multiverse of bubble universes of varied physical “landscape”, plus observers in other parts of the quantum superposition of wave functions for that multiverse, and perhaps even observers in other mathematically possible universes \cite{tegmark}, all aggregated over infinite time, and that gives a potentially very large reference class. However, if we use the same large reference class in both frameworks, coupling it on the one hand to the theory $T$ of early convergence in a relatively “little neighborhood” of the multiverse, and on the other hand to the theory $\T$ of late or no convergence in this neighborhood, then the salient effects of the larger reference class would seem to cancel out, being just a repetition of similar situations in other “little neighborhoods” of the vast multiverse, leaving us again with the same results. 

So the result appears to be robust with respect to reference class. Thus, while we don't have answers about the future fate of life, we do have hints, and these hints suggest there may be something extraordinary to come. 

\section{Acknowledgements}
I'd like Anthony Aguirre, Federico Faggin, Yuri Milner, Doug Abrams, Ed Schulak, and Don Antonel for stimulating discussions. 


\end{document}